# Seeds Buffering for Information Spreading Processes

Jarosław Jankowski[1], Piotr Bródka[2],
Radosław Michalski[2], and Przemysław Kazienko[2]

[1] Faculty of Computer Science for Information Technology
West Pomeranian University of Technology
Szczecin, Poland
jjankowski@wi.zut.edu.pl
[2] Department of Computational Intelligence,
Wrocław University of Science and Technology, Wrocław, Poland
{piotr.brodka,radoslaw.michalski,kazienko}@pwr.edu.pl

**Abstract.** Seeding strategies for influence maximization in social networks have been studied for more than a decade. They have mainly relied on the activation of all resources (seeds) simultaneously in the beginning; yet, it has been shown that sequential seeding strategies are commonly better. This research focuses on studying sequential seeding with buffering, which is an extension to basic sequential seeding concept. The proposed method avoids choosing nodes that will be activated through the natural diffusion process, which is leading to better use of the budget for activating seed nodes in the social influence process. This approach was compared with sequential seeding without buffering and single stage seeding. The results on both real and artificial social networks confirm that the buffer-based consecutive seeding is a good trade-off between the final coverage and the time to reach it. It performs significantly better than its rivals for a fixed budget. The gain is obtained by dynamic rankings and the ability to detect network areas with nodes that are not yet activated and have high potential of activating their   neighbours.

## 1  Introduction

The growing complexity of problems which need to be solved daily is leading to increasingly complicated decision processes. In order to reduce risk and uncertainty, some decisions are naturally divided into a sequence of less complicated component decisions. Even though the decision can be taken and implemented immediately, [9, 19], it is not always the most efficient strategy from the perspective of the final outcome, especially assuming that the process at hand bears some uncertainty. As an alternative, sequential analysis and decisions where introduced by Wald [31] and extended later [29, 3]. Lower risk is usually assigned to series of many smaller decisions. As an outcome, dividing the decision or activity into smaller chunks can be more profitable compared to the decisions that are taken immediately. In terms of acquiring knowledge and reducing risk, instead of using partial knowledge in the first stage, the strategy that performs better in many cases is gathering more knowledge about the nature of the process during its runtime, and using that knowledge in future decision-making. The same applies to marketing [30], cognitive science [14], medicine, especially for vaccination strategies [26, 13], and the sequential nature of the process is also found in nature; for instance, in the way viral infections develop [4].

In this project, we investigate two seeding strategies built upon sequential seeding, which were initially proposed in [7]: (i) *sequential seeding with revival* and (ii) *sequential seeding with buffering*. both are suited for the social influence maximisation problem [10], which was extended in various directions [21], [18]. They are based on an independent cascades model and the concept of dynamic seed allocation, in which seeds are not used until the natural diffusion process stops. Yet, they differ on how the seeds are activated after the diffusion terminates. The proposed strategies are compared with typical single stage seeding when all seeds are used in the first stage, as in most typical seeding strategies [9, 19, 10, 11]. The main goal of the work is to verify the performance of the proposed approach for different parameters related to network structures and characteristics of diffusion processes. The initial research on sequential seeding showed that the same number of seeds activated over time offers better results, i.e. a larger final coverage, compared to the single stage seeding [7], so the natural research question is whether there is any chance to outperform it by introducing some novel features. In this study, we carried out detailed research on better understanding this phenomenon, with the verification of several strategies that expand a typical sequential seeding strategy, which is using many parameters of diffusion processes.

## 2   Conceptual Framework

Earlier research showed a better outcome of the sequential seeding, compared to a single stage seeding, due to better usage of potential of natural diffusion processes. In this section, we further discuss methods for better exploitation of sequential seeding. Results from the single stage seeding (*SS*) are treated as a reference for evaluation of performance of sequential seeding. In the first stage, *n* initial seeds are selected with the use of a seed selection strategy, e.g., based on the structural characteristics of nodes like the degree, closeness, etc. . The diffusion process starts and continues without any additional support, until it naturally stops at the time $T_{SS}$, (see Fig. 1. Its coverage is measured by a percentage of the naturally activated nodes in relation to all nodes in the network, and represented by $C_{SS}$. In the proposed generic sequential approach, splitting the seeds among several stages takes place in a form of sequence of seeds, which are used in several consecutive stages of the process [7]. The selection of seeds in each step is based on the ranks built using static measures, which were computed once before the process begins. The next evaluated approach is based on the highest decomposition of the seeding budget over time, and by activating a single node per stage (*OP S_Sq*).

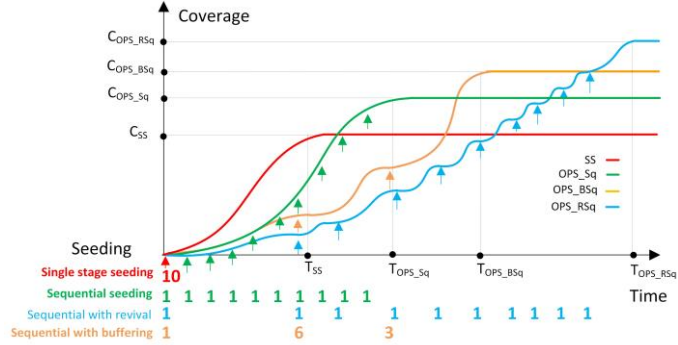

Fig. 1: Macroscopic view of diffusion process with various seeding strategies applied: single stage seeding *SS*, sequential seeding with one seed per stage activated *OP S_Sq*, sequential seeding with revival *OP S_RSq* and with buffering *OP S_BSq*. Only natural activations are contributing to the coverage.

The *OP S_Sq* approach is not dependent on the process characteristics and it leaves room for improvement, because the allocation of additional seeds takes place, even when natural diffusion processes are characterised with high dynamics. In order to use acquired knowledge about this dynamics, a sequential strategy with a revival mode is proposed, and this is named *OP S_RSq*. Additional seeding is used only when the diffusion process stops. In this approach, when natural activations are observed, additional seeds are not used. The proposed approach extends the total period of seeding because the number of seeds used to improve dynamics is the same as the ones present in generic strategies, and additional seeding is postponed to other periods if no further natural activations are detected. Here, only one seed is used per stage. Due to specifics of an independent cascades model, each node has only one chance to activate neighbours. Then, if the natural process stops, it will not be re-initiated in a natural way during future stages, so the only way to continue activations is by using additional seeding.

The second proposed method that extends the capabilities of sequential seeding is *sequential seeding with buffering - OP S_BSq*. Here, unlike with sequential seeding, if the activation process unfolds naturally and no additional seeding is performed. Yet, i for each stage that didnt require any seeding, a virtual counter increases the value of the amount of seeds which will be activated after the natural activation plateau is reached. If so, the number of seeds *n* which is equal to the value of this counter is activated and the counter is reset. All four seeding methods are depicted in Fig. 1 and the approaches which were based on sequential seeding are summarized in Table 1.

Table 1: Sequential seeding strategies

| | |
|---|---|
| **OPS Sq** | one per stage generic sequential seeding based on atomic decomposition with the one per stage seed used in each step of simulations and the length of sequence equals the number of seeds used |
| **OPS RSq** | one per stage seeding with revival mode; an additional seed is used only when the natural diffusion process finishes |
| **OPS BSq** | one per stage seeding with buffering mode where additional seeds are collected in the buffer while natural processes continues; the seeds from the buffer are used after the natural diffusion process stops |

## 3   Related Work

The original influence maximization problem [10] considered static social networks, and researchers followed that path when proposing new algorithms for tackling it. Moreover, they focused on single stage seeding, i.e., the best allocation of the budget assuming its immediate spending [20] without any further support. However, in many realistic scenarios, this is not the only way of managing the budget and recent studies started to investigate how distributing a budget over time influences the outcome of the process. Starting with two-stage stochastic models [27] through more scalable approaches [6], it was shown that this research direction has a potential that was further explored in the area of social influence [28, 34]. The most recent study [7] proposed sequential seeding as being a more effective way to allocate the budget, while showing its advantages over a single stage seeding in many network configurations. The potential of multi-period spraying algorithm for routing in delay-tolerant networks was also discussed [2]. Apart from that, some authors noticed the benefit of omitting the nodes that will be activated due to the natural diffusion process. As a result they proposed methods that avoid choosing as a new seed either a node that is a neighbour of the already chosen seed, [12, 35], or a node that is in a local cluster that already contains another seed, which also reflects the observation on how information spreads across clusters [8]. In this assignment, we investigate sequential seeding [7] in detail, so as to answer the question about the extent to which this approach provides a better solution, compared to a single stage seeding.

## 4   Experimental Setting

Experimental research was conducted, using agent-based simulations within twenty networks, and including eleven real networks (N1 [22], N2 [22], N3 [24], N4 [22], N5 [1], N6 [33], N7 [23], N8 [25], N9 [17], N10 [16], N11 [15]) and nine artificially generated networks following the Watts-Strogatz and Barabsi-Albert models (N12-N20); these were done according to the specifications presented in Table 2. Networks were generated with the use of various parameters. For Watts-Strogats model Parameter 1 represents the neighbourhood within which the vertices of the lattice will be connected and Parameter 2 represents rewiring probability. In the case of the Barabasi-Albert model, Parameter 1 represents the number of edges to be included in each time step and Parameter 2 shows the power of the preferential attachment.

Table 2: Specification of synthetic networks N12-N20

| ID  | Network model   | Param. 1 | Param. 2 | Nodes  | Edges   |
|-----|-----------------|----------|----------|--------|---------|
| N12 | Watts-Strogatz  | nei=1    | r=0.05   | 10,000 | 20,000  |
| N13 | Watts-Strogatz  | nei=2    | r=0.05   | 10,000 | 60,000  |
| N14 | Watts-Strogatz  | nei=3    | r=0.05   | 10,000 | 120,000 |
| N15 | Watts-Strogatz  | nei=2    | r=0.10   | 10,000 | 60,000  |
| N16 | Watts-Strogatz  | nei=2    | r=0.30   | 10,000 | 60,000  |
| N17 | Watts-Strogatz  | nei=2    | r=0.50   | 10,000 | 60,000  |
| N18 | Barabasi-Albert | m=2      | p=0.50   | 10,000 | 19,997  |
| N19 | Barabasi-Albert | m=4      | p=0.50   | 10,000 | 39,990  |
| N20 | Barabasi-Albert | m=8      | p=0.50   | 10,000 | 79,964  |

The independent cascades model (IC) [10] was used for each edge ($a, b$), with the propagation probability $PP(a, b)$ that node $a$ activates node $b$ in the step $t + 1$, with the condition that node $a$ was activated in the time $t$ [32].

The main reason for the selection of IC model is that in the IC model, a single seed can induce diffusion and even a cascade, while in the linear threshold model [5], small seeds packages would not have any effect.

## 5 Results

### 5.1 Sequential seeding with revival and buffering

Results achieved in sequential seeding were compared to the single stage seeding (SS) in the same conditions, i.e., for the same network and its parameters, such as propagation probability, seeding percentage and seed selection strategy based on a random selection (R), the degree (D), the second-level degree (D2), the closeness (CL), the clustering coefficient (CC) or the PageRank (PR) across twenty different networks. Reference values for comparison are based on the coverage achieved for single stage seeding ($C_{SS}$) and the duration of the single stage process representing the stage when the $T_{SS}$ is achieved. Experiments showed that sequential seeding was almost always better than single stage seeding with the same parameters. An example of s simulation case is presented in Fig. 2.

Results for all networks, strategies and parameters showed that, in 91.94% of simulation cases, $OPS\_Sq$ delivered better results than single stage seeding.

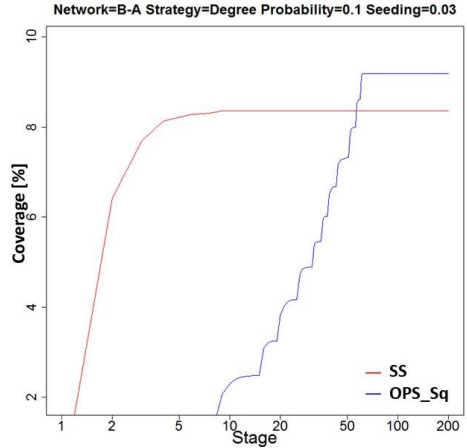

Fig. 2: Longer duration of sequential seeding with higher coverage

Even though the performance was dependent on the network characteristics and parameters of the process and the strategy used, the sequential seeding supported the diffusion in most cases. The improvement can even exceed 50% with the use of the same number of seeds, just like in single stage seeding. An average reach of diffusion processes based on the *OP S Sq* with statistical significance ($p < 2.2e^{-16}$) achieved 8.43% better results than using the *SS* approach with the same conditions based on Wilcoxon signed-rank test.

The analysis of *OP S Sq* showed that sequential seeding outperforms single stage seeding in most cases, and the performance of proposed methods is dependent on parameters of the diffusion process and network characteristics. Experiments were performed for a wide range of networks and parameters, including a very low performance, i.e., the propagation probability PP=0.01 or seeding percentage 1% with a low number of activations and very difficult diffusion processes, no matter what strategy is applied. An opposite situation is observed within the networks, with a high degree and propagation probability PP=0.25 and SP=5%. Under these conditions, most strategies were performing very well, with diffusion processes leading to 100% activated nodes in a very short time, and they left a very small margin for improvements. Taking into account the above conditions, an average 8.43% or 6% of increase shows substantial growth, with much better results for conditions such as a higher activation probability. A substantial increase was obtained for degree -based strategies for both one (D) and second level degree (D2), and a propagation probability higher than 0.05.

For networks N2, N4, N10 and N19, a high performance average reaching higher 30% than in single stage seeding was observed for both sequential strategies, based on D and D2 selection, see Table 3 containing results compared with single stage seeding for each network. Low performance was observed for networks N12, N14, and at least in three strategies for networks N6, N8; however, it was still above 10% with only four cases with an increase which was smaller than 5% (N12, N14).

The first reason why sequential seeding (including its modifications outperforms single stage seeding is the fact that, in the case of sequential seeding, initial seeds used in a single stage approach are activated by a natural process, due to their high positions in the network, and they dont require seeding to be activated. When taking into account the seeds selected in the single stage seeding process, more than 60% of them can be activated in natural processes when sequential seeding is applied. Saved seeds can be used for activation of other nodes and unexplored segments of the network. This refers mostly to revival mode strategies, where the phenomena based on using natural diffusion processes is most visible, with additional seeding performed only when natural activations are stopped.

Table 3: Results for $OP\_S\_Sq$ based on D and D2 strategies with PP=0.1 and SP=0.05

| Strategy | N1 | N2 | N3 | N4 | N5 |
|---|---|---|---|---|---|
| OPS_Sq D | 124.81 | 138.66 | 111.82 | 135.87 | 110.36 |
| OPS_Sq D2 | 120.66 | 149.65 | 113.06 | 145.35 | 109.99 |
| Strategy | N6 | N7 | N8 | N9 | N10 |
| OPS_Sq D | 107.47 | 126.21 | 113.05 | 109.05 | 130.22 |
| OPS_Sq D2 | 113.89 | 137.28 | 111.33 | 114.01 | 145.41 |
| Strategy | N11 | N12 | N13 | N14 | N15 |
| OPS_Sq D | 112.29 | 102.94 | 114.67 | 106.45 | 116.16 |
| OPS_Sq D2 | 112.49 | 106.15 | 115.37 | 106.15 | 117.69 |
| Strategy | N16 | N17 | N18 | N19 | N20 |
| OPS_Sq D | 119.59 | 114.94 | 114.44 | 130.78 | 114.28 |
| OPS_Sq D2 | 118.84 | 119.24 | 119.33 | 130.59 | 114.60 |

In the next stage, an approach based on seeding with revival mode (with the presence of additional seeding) was used, and diffusion processes stops were observed. Results for compared strategies with the revival model showed statistical differences between $OP\_S\_Sq$ vs $OP\_S\_RSq$ with $p < 2.2e-16$. Results showed that the revival mode $OPS\_RSq$ achieved a 5.34% better reach than $OPS\_Sq$ for all cases linked to static rankings. In 2319 cases (64.42%), $OPS\_RSq$ delivered better results than $OP\_S\_Sq$. For $OP\_S\_RSq$, the best performance above median 4.66% was observed for networks N4, N5, N7, N8, N9, N10, N13, N14, N16, and N20; when compared to the non-revival mode, network parameters were not statistically significant. The results showed that, for networks with a higher degree, the performance of the revival mode was better. Diffusion processes within networks with a higher degree have higher dynamics and allocation of additional seeds when natural processes are continued is resulting in the waste of available resources. Results analysed for seeding percentages 1%, 2%, 3%, 4%, 5% showed an increase of *OPS_RSq* with values of 11.79%, 7.42%, 3.42%, 3.40%, 2.30%, and the relation between the seeding performance and revival mode showed the highest increase of performance for small proportion of seeds selected ($SP = 1\%$). While most of the analysis is performed on aggregated data obtained from all cases, Fig. (3 shows example results from a simulation performed within the network N2, with the propagation probability PP=0.1 and seed selection strategy D2, and 5% of the initial selected seeds s. The revival mode is achieving better results, but the process is longer and dynamics are smaller than what is observed from the 75*th* stage of simulation.

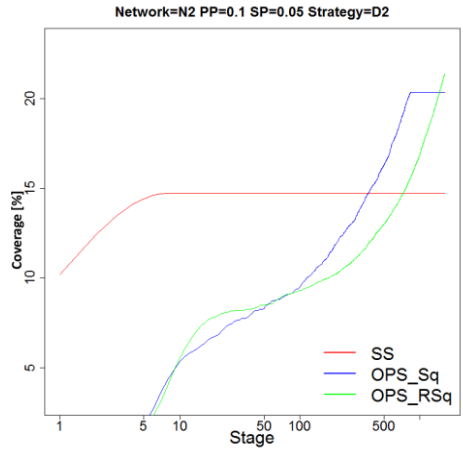

Fig. 3: Revival mode results within network N2 with PP=0.1 and SP=5% based on S=D2

The introduced approach to the revival mode increased the reach of processes, but the duration was increased as well, due to the delay of each additional seeding, until diffusion processes stopped. Results for compared strategies with the Wilcoxon signed rank and revival mode showed statistical differences between *OPS_Sq* vs. *OPS_RSq* and $p = 4.40e^{-09}$). The buffered mode *OPS_BSq* achieved 3.2% increase of reach when compared to *OPS_Sq*, and it represented 60% of value achieved with the revival mode (5.34%). The buffering mode, when compared to the revival mode without buffering, makes it possible to shorten the duration of processes achieved better results than in the case of generic sequential seeding. The average duration of *OPS_BSq* was only 21.92% longer than *OPS_Sq*, while the duration of *OPS_RSq* was 71.31% longer. Fig. 4a illustrates the duration of *OPS_BSq* and *OPS_RSq* in relation to generic *OPS_Sq* approach.

The buffered approach showed an average increase when compared to *OPS Sq* for propagation probabilities 0.01, 0.05, 0.1, 0.15, 0.2, 0.25 as follows: 8.48%, 4.24%, 3.22%, 4.21%, 2.54% and 0.00%, with the best result being for lowest propagation probabilities. When compared to the revival mode, the buffering mode delivered improvements for PP=0.01 with a 2.61% increase, while for probabilities 0.05, 0.1, 0.15, 0.2, 0.25, buffering results were worse. Average results from all propagation probabilities showed slightly better results than in the revival mode $OPS\_RSq$ for 6 networks: N1, N6, N12, N16, N17 and N15. Fig. 4b shows an example $OPS\_BSq$ and $OPS\_RSq$ results from network N10 with the PP=0.1 strategy D2 and SP=0.05 with higher reach achieved for $OPS\_RSq$ and a lower reach of $OPS\_BSq$, but with a shorter duration proposes.

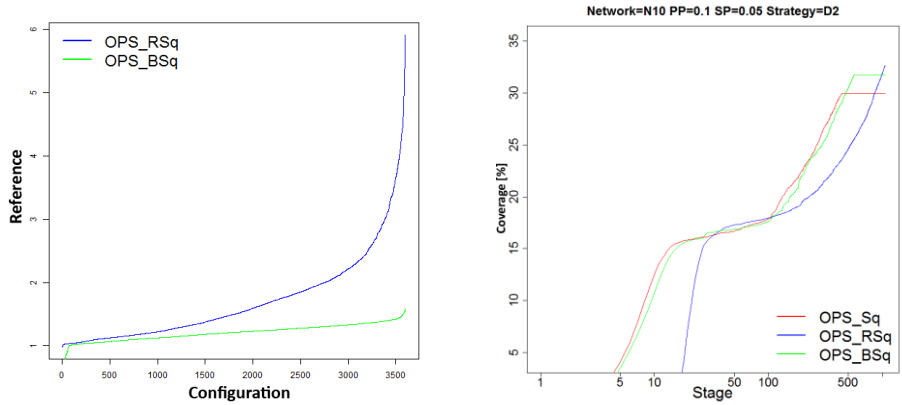

(a) Duration of difussion processes for $OPS\_BSq$ and $OPS\_RSq$ compared with generic $OPS\_Sq$

(b) Strategies $OPS\_BSq$ and $OPS\_RSq$ compared with sequential seeding for network N10 with $PP = 0.1$, strategy D2 and $SP = 0.05$

Fig. 4: Comparison of seeding strategies

Sequential strategies delivered better results than the single stage approach in terms of reach; however, the longer duration of diffusion processes is a disadvantage for the higher reach. It is a result of the smaller amount of seeds used in the first stages of processes and lower initial dynamics, due to spreading seeds over several periods of time. In an analysis presented in this section, the stage $T_{SS}$ was used as a reference, where the maximal number of activations is achieved using a single stage approach. The cases in Fig. 5 show differences among generics per stage seeding strategy, as well as its extensions with the revival and buffering modes. The duration of one per stage seeding was increased by the revival mode, but cases from the buffering mode are concentrated between $OPS\_Sq$ and $OPS\_RSq$.

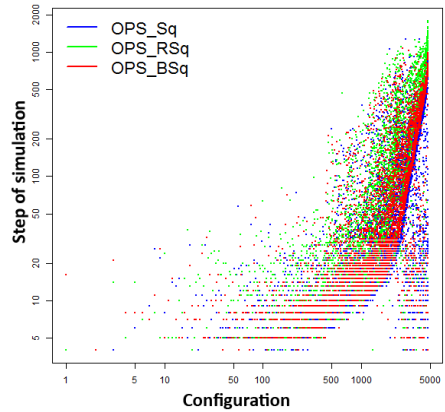

Fig. 5: Duration of one per stage strategies *OP S Sq*, *OP S RSq* and *OP S BSq*

## 5.2  Trade-off between the coverage and duration of the process

The potential of sequential seeding can be evaluated as a trade-off between coverage and duration. The last stage of analysis includes connected effects of increased coverage and longer duration of the diffusion processes with the reference to a single stage seeding, with integrated results for the same PP, SP, N and S. The distribution of reference values for *OP S Sq* for all used strategies shows visual characteristics of obtained results in Fig. 6. The X axis represents the distance from the single stage reach, which was computed with the formula *Coverage$_{Ref}$* = (*Coverage$_{Sq}$* − *Coverage$_{SS}$*)/*Coverage$_{SS}$*. Negative values represent cases with worse results than in a single stage seeding. Here, the Y axis is representing the distance between the stage when a maximal value was achieved according to the formula *Duration Ref* = (*Duration Sq* − *Duration SS*)/*Duration SS*. Each figure includes 3,600 cases for each sequential strategy. References for *OP S Sq* are characterized by a bigger dispersion on the Y axis, including more cases with high values which are achieved in the X axis.

The effect of the revival mode for *OP S RSq* is presented in Fig. 7a with a growing dispersion and longer sequences, but with a shift that was observed on the X axis, and geared towards better reach. The reach and duration of diffusion processes and the relation between them in reference to single stage seeding were connected with results of a simulation with the same parameters. Cases with negative values on the X axis represent worse results in terms of reach than with single stage seeding. Cases with negative values on the Y-axis represent processes with *SS SS*, which are achieved faster than in a single stage approach. Results for the buffered mode are presented in Fig. 7b. Buffered mode resulted in dispersion between *OP S Sq* and *OP S RSq* in terms of X and Y distribution.

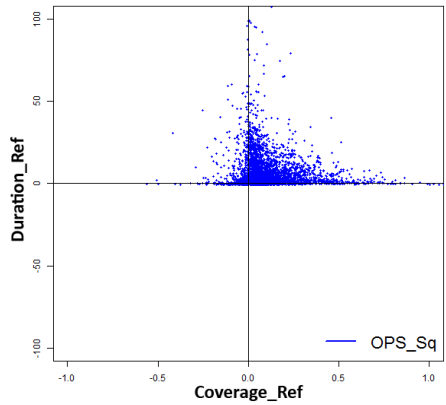

Fig. 6: Distribution of cases in generic sequential seeding approach *OP S Sq*

Distribution of cases with used *OPS_Sq* shows a bigger dispersion and longer duration, especially for cases with low X values. Implementation of revival mode resulted in an increased number of cases, with a longer duration and an accompanied increase of reach.

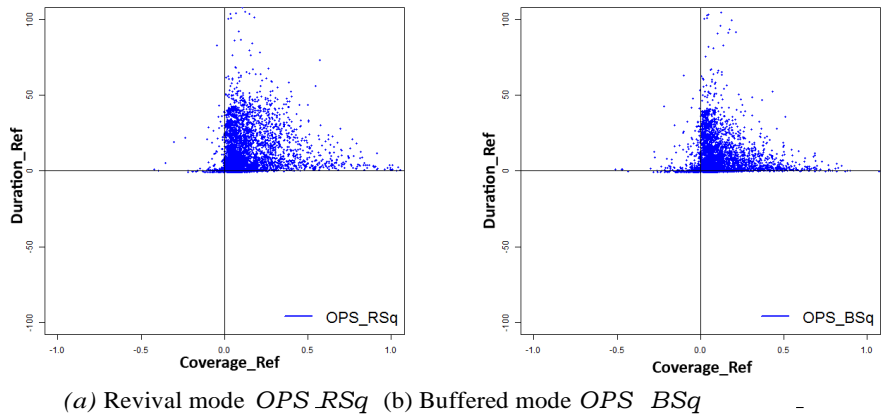

*(a)* Revival mode  *OPS_RSq*   (b) Buffered mode *OPS   BSq*

Fig. 7: Distribution of cases

## 6   Conclusions

The results revealed that the best result can be achieved by means of one per stage method with the revival mode, and it also lasts the longest. This observation is valid for all tested ranking methods: degree, second level degree, closeness, clustering coefficient and page rank If the process time span is limited, even from the duration of the single stage approach (the shortest possible), the length of the sequence may have to be reduced and the buffer-based consecutive seeding appears to be a good solution. The results show the potential of the buffering mode as being a suitable compromise, which enables us to extend coverage without high increase in duration. During the experiments, results from short sequences were compared with the longest sequences based on revival concepts. The sequential seeding approaches with the longer duration usually yield better results in terms of the number of activated nodes, which is the final coverage. Hence, the trade-off between time and reach depends on individual preferences and the role of process coverage and duration in a given application domain. Further research will focus on various modifications of the buffering mode, as well as on other approaches for duration shortening. The general direction is leading towards finding clear dependencies between the length of sequences and the final coverage for different networks and strategies.

## Acknowledgments

This work was partially supported by Wrocław University of Science and Technology statutory funds (PB) and by the National Science Centre, Poland, project no. 2015/17/D/ST6/04046 (RM), 2016/21/B/ST6/01463 (PK), 2013/09/B/ST6/02317 and 2016/21/B/HS4/01562 (JJ); by European Union's Horizon 2020 research and innovation programme under the Marie Skłodowska-Curie grant no. 691152 (RENOIR); by the Polish Ministry of Science and Higher Education fund for supporting internationally co-financed projects in 2016-2019, no. 3628/H2020/2016/2.